# Hamiltonian approach to slip-stacking dynamics

S. Y. Lee

*Department of Physics, Indiana University, Bloomington, Indiana 47405, USA*

K. Y. Ng

*Fermi National Accelerator Laboratory, Batavia, Illinois 60510, USA*

Hamiltonian dynamics has been applied to study the slip-stacking dynamics. The canonical-perturbation method is employed to obtain the second-harmonic correction term in the slip-stacking Hamiltonian. The Hamiltonian approach provides a clear optimal method for choosing the slip-stacking parameter and improving stacking efficiency. The dynamics are applied specifically to the Fermilab Booster-Recycler complex. The dynamics can also be applied to other accelerator complexes.

## I. INTRODUCTION

Beam intensity is limited by space-charge effects at low energies. Rapid cycling synchrotrons (RCS) can be used to increase beam power. However, RCS is usually limited by its achievable energy and a second-stage accelerator is required to increase both energy and beam power. Slip-stacking injection may be a necessary method to achieve high power through doubling the beam intensity. The idea of slip-stacking was first proposed by Mills [1], where he studied the stability of particle motion under the influence of an rf system at a nearby frequency. An experimental trial on slip-stacking had been carried out in the CERN SPS (European Organization for Nuclear Research, Superproton Synchrotron) to accumulate beams from the CERN PS (Proton Synchrotron) [2]. Slip-stacking had later been applied in the Fermilab Tevatron Run IIB, where proton bunches from the Booster were slip-stacked in the Main Injector in order to increase antiproton production, when the slip-stacked proton beam was extracted to hit a target [3]. At the moment, slip-stacking is also employed in the Fermilab Recycler Ring to increase the proton beam power for neutrino production [4].

Recently, the dynamics of slip-stacking have been studied by Eldred [5], by solving single-particle equations of motion through numerical simulations. Eldred also proposed second-harmonic correction to the slip-stacking rf system. Here, we wish to attack the same problem but in the Hamiltonian approach. This approach appears to be much simpler to provide the second-harmonic rf voltage correction required to cancel the interaction between the two series of slipping rf buckets (or resonance islands). In addition, it also provides us with the understanding of the parametric resonances that pop up in the rf buckets and the formation of chaotic regions during the slip-stacking operation. Our paper is organized as follows. In Sec. II, we lay out the Hamiltonian for the slip-stacking equation of motion, where the slip-stacking Hamiltonian depends only on a single slip-stacking parameter $\alpha_s$. In Sec. III, we carry out canonical perturbation to recover the resonances produced by mutual interaction between the upper and lower buckets. It becomes clear that these resonances can be canceled by using an additional second-harmonic rf system. In Sec. IV, we carry out numerical simulations to verify the effects attained from the Hamiltonian dynamics. In Sec. V, we discuss the choice of the slip-stacking parameter and apply it specifically to the Fermilab accelerator complex. Conclusion of our work is discussed in Sec. VI.

## II. THE SLIP-STACKING HAMILTONIAN

When there is only one rf system in a synchrotron, the Hamiltonian describing the longitudinal motion is [6]

$$H = \frac{\nu_s p^2}{2} + \nu_s [1 - \cos\phi], \qquad (2.1)$$

where, for the sake of simplicity, we use the *normalized* off-momentum coordinate $p = h|\eta|\delta/\nu_s$ with $\delta \equiv \Delta P/P_0$ being the nominal fractional off-momentum of a particle, $P$ the momentum of the particle and $P_0$ the nominal momentum of the beam, $h$ the harmonic number of the rf system, and $\eta$ the phase-slip factor of the slip-stacking ring. The small-amplitude synchrotron tune is $\nu_s = \sqrt{h|\eta|eV_{\rm rf}/2\pi\beta^2 E}$, where $V_{\rm rf}$ is the rf voltage, and $\beta c$ and $E$ are the nominal speed and energy of the beam particles. The rf phase $\phi$ and the normalized off-momentum $p$ are conjugate canonical coordinates, while $\theta$ represents the independent "time" variable, which increases by $2\pi$ in This manuscript has been authored by Fermi Research Alliance, LLC under Contract No. DE-AC02-07CH11359 with the U.S. Department of Energy, Office of Science, Office of High Energy Physics. The U.S. Government retains and the publisher, by accepting the article for publication, acknowledges that the U.S. Government retains a non-exclusive, paid-up, irrevocable, world-wide license to publish or reproduce the published form of this manuscript, or allow others to do so, for U.S. Government purposes.

each revolution around the ring. In the normalized phase-space coordinates $p$ and $\phi$, the rf bucket is pendulumlike in shape and is bounded by $|\phi| \leq \pi$ and $|p| \leq 2$. We call, in this paper, this rf bucket the *unperturbed rf bucket*, meaning that it is not influenced by any other rf system. Every point inside the unperturbed rf bucket is stable.

During slip-stacking, there is another rf system at an rf frequency lower or higher than the fundamental one by $f_{slip}$. In order for slip-stacking to work, these two rf systems must generate buckets which slip by exactly one train or one batch of rf buckets in consecutive injections from the rapid-cycling booster synchrotron. This condition fixes the rf frequency difference to $f_{slip} = h_B f_B$, where $f_B$ is the repetition frequency of the RCS and $h_B$ is its rf harmonics. For Fermilab, the RCS is the Booster with $h_B = 84$ and $f_B = 15$ Hz. The harmonic number of the Recycler ring is $h_R = 588$. Hereafter, all symbols without subscript represent parameters of the slip-stacking ring. For convenience, we introduce a *slip-tune* $\nu_{slip} = f_{slip}/f_0$, where $f_0$ is the revolution frequency of the Recycler Ring.

In the presence of the two rf systems, the Hamiltonian becomes [6]

$$H = \frac{\nu_s p^2}{2} + \nu_s [2 - \cos\phi - \cos(\phi - \nu_{slip}\theta)]. \quad (2.2)$$

Here, a constant 2 is added to the Hamiltonian for convenience, and we assume that these two rf systems have the same total rf cavity voltage $V_{rf}$. Note that the rf buckets generated by the rf system corresponding to $\cos(\phi - \nu_{slip}\theta)$ are at a slightly lower energy than the buckets generated by the fundamental rf system corresponding to $\cos\phi$ in the Hamiltonian (if the phase-slip factor is $\eta < 0$). Because of the lower energy, the lower-energy bucket series slips forward at the rate of $\Delta\phi = \nu_{slip}\theta$. The fractional momentum that separates the upper and lower bucket series is

$$\Delta\delta_{sep} \equiv \frac{\Delta P_{sep}}{P_0} = \frac{\nu_{slip}}{h|\eta|} = \frac{h_B f_B}{h_R |\eta| f_0}, \quad (2.3)$$

where $\Delta P_{sep}$ is the momentum difference of the two slip-stacking beams, $P_0$ is the nominal momentum of the beams, and $\eta$ is the phase slip-factor of the slip-stacking ring. Once the repetition rate of the RCS, the phase-slip factors, and the revolution frequency of the slip-stacking ring are designed, the momentum separation of the two bucket series, $\Delta\delta_{sep}$, is fixed. In terms of the normalized off-momentum coordinate $p$, the separation of the centers of the upper and lower buckets is

$$\Delta p_{sep} = \frac{h|\eta|\Delta\delta_{sep}}{\nu_s} = \frac{\nu_{slip}}{\nu_s} \equiv \alpha_s, \quad (2.4)$$

where $\alpha_s$ is called the *slip-stacking parameter* with the property $\alpha_s \propto V_{rf}^{-1/2}$. The unperturbed bucket height is

$|p| \leq 2$, which is independent of $\alpha_s$. Therefore the unperturbed rf buckets of the two rf systems just touch each other at $\alpha_s = 4$ [1]. The two unperturbed rf buckets are separated from each other when $\alpha_s > 4$, and they overlap when $\alpha_s < 4$.

Because of the presence of the two rf systems, the two series of rf buckets, upper and lower, are mutually perturbing each other. The stable bucket areas become smaller than those of the unperturbed rf buckets. When the upper and lower series of buckets overlap, resonance islands can be generated in between the two series of rf buckets usually around $p = 0$ and $\phi = 0$ and $\pm\pi$. In addition, chaotic regions may be created, which can reduce the stable region of the rf buckets significantly.

Overlapping resonances can be avoided if the upper and lower buckets are widely separated or if $\alpha_s \gg 4$. Bigger $\alpha_s$, however, implies smaller rf voltage and therefore smaller unperturbed bucket areas (in $\delta - \phi$ coordinates), which may not be large enough to accommodate the beam injected from the RCS. On the other hand, smaller $\alpha_s$ implies larger rf voltage. One may think that there would be bigger unperturbed bucket areas to accept the beam injected from the RCS. When $\alpha_s < 4$, these two bucket series can produce strong overlapping resonances and chaos so that the stable parts of the buckets become smaller than the unperturbed buckets.

We now examine slip-stacking in the *unnormalized* off-momentum coordinate $\delta = \Delta P/P_0$. The bucket separation $\Delta\delta_{sep}$ given by Eq. (2.3) is rf-voltage independent, while the half-momentum widths of the unperturbed buckets (or half-bucket height)

$$\Delta\delta_{bucket} = \sqrt{\frac{4eV_{rf}}{2\pi\beta^2 Eh|\eta|}} = \frac{2\nu_s}{h|\eta|} \quad (2.5)$$

are proportional to $V_{rf}^{1/2}$. When $\Delta\delta_{sep} = 2\Delta\delta_{bucket}$, the unperturbed upper rf bucket touches the unperturbed lower rf bucket, which again reduces to $\alpha_s = 4$. As rf voltage $V_{rf}$ is reduced so that $\Delta\delta_{sep} \gg 2\Delta\delta_{bucket}$ (or $\alpha_s \gg 4$), the unperturbed bucket height $\Delta\delta_{bucket}$ decreases and so is the bucket area. The bucket may not be big enough to accommodate the injected bunch. The two series of rf buckets, however, are father apart. However, if $V_{rf}$ is made larger so that $\Delta\delta_{sep} < 2\Delta\delta_{bucket}$ (or $\alpha_s < 4$), the upper and lower bucket series overlap each other. This generates resonance islands and chaotic regions in phase space, which in turn will reduce the stable areas of the upper and lower buckets.

The interaction of these two rf systems will produce resonances in between the upper and lower buckets. If one can find a compensation rf system to cancel the driving term, these additional resonances can be eliminated or minimized, and one may be able to restore the slip-stacking buckets for the injection bunches.

To simplify the derivation, we first symmetrize the Hamiltonian to a frame with the upper and lower buckets centered at $p = +\frac{1}{2}\alpha_s$, or with frames moving at $\frac{1}{2}\nu_{\text{slip}}\theta$. This can be accomplished by a canonical transformation using the generating function

$$F_2(\phi, \tilde{p}) = \left(\phi - \frac{\nu_{\text{slip}}\theta}{2}\right)\left(\tilde{p} + \frac{\alpha_s}{2}\right), \qquad (2.6)$$

where the new and old canonical variables are related by

$$\tilde{\phi} = \frac{\partial F_2}{\partial \tilde{p}} = \phi - \frac{\nu_{\text{slip}}\theta}{2} \quad \text{and} \quad p = \frac{\partial F_2}{\partial \phi} = \tilde{p} + \frac{\alpha_s}{2}. \quad (2.7)$$

The new Hamiltonian is

$$H = \frac{\nu_s \tilde{p}^2}{2} + \nu_s\left[2 - \cos\left(\tilde{\phi} + \frac{\nu_{\text{slip}}\theta}{2}\right) - \cos\left(\tilde{\phi} - \frac{\nu_{\text{slip}}\theta}{2}\right)\right] + \frac{\nu_s \alpha_s^2}{8}, \qquad (2.8)$$

where $\tilde{p}$ and $\tilde{\phi}$ are the conjugate canonical coordinates of the symmetrized slip-stacking rf systems. The Hamiltonian represents the upper and lower buckets moving at $\Delta\phi = \mp\nu_{\text{slip}}\theta/2$, respectively, while the structures in between the two buckets centered at $\tilde{p} = 0$ is stationary. We will focus on the phase space near $\tilde{p} = 0$.

## III. THE SECOND-ORDER CANONICAL PERTURBATION

Ignoring the uninteresting constant term in Eq. (2.8) and simplifying notation by reidentifying phase-space coordinates with $\tilde{p} \to p$ and $\tilde{\phi} \to \phi$, the Hamiltonian is

$$H = \frac{\nu_s p^2}{2} + \nu_s\left[2 - \cos\left(\phi + \frac{\nu_{\text{slip}}\theta}{2}\right) - \cos\left(\phi - \frac{\nu_{\text{slip}}\theta}{2}\right)\right]. \qquad (3.1)$$

Because we wish to study the phase space structure in between the upper and lower rf buckets, we perform a canonical transformation to cancel the potential energy part of the Hamiltonian (3.1) using the generating function

$$F_2(\phi, \bar{p}) = \phi\bar{p} + a(\bar{p})\sin\left(\phi + \frac{\nu_{\text{slip}}}{2}\theta\right) + b(\bar{p})\sin\left(\phi - \frac{\nu_{\text{slip}}}{2}\theta\right), \qquad (3.2)$$

where $a(\bar{p})$ and $b(\bar{p})$ are two functions of $\bar{p}$ to be determined. The transformation is

$$\bar{\phi} = \frac{\partial F_2}{\partial \bar{p}} = \phi + a'(\bar{p})\sin\left(\phi + \frac{\nu_{\text{slip}}}{2}\theta\right) + b'(\bar{p})\sin\left(\phi - \frac{\nu_{\text{slip}}}{2}\theta\right), \qquad (3.3)$$

$$p = \frac{\partial F_2}{\partial \phi} = \bar{p} + a(\bar{p})\cos\left(\phi + \frac{\nu_{\text{slip}}}{2}\theta\right) + b(\bar{p})\cos\left(\phi - \frac{\nu_{\text{slip}}}{2}\theta\right). \qquad (3.4)$$

The transformed Hamiltonian is

$$H = \frac{\nu_s}{2}\left[\bar{p} + a(\bar{p})\cos\left(\phi + \frac{\nu_{\text{slip}}}{2}\theta\right) + b(\bar{p})\cos\left(\phi - \frac{\nu_{\text{slip}}}{2}\theta\right)\right]^2 + \nu_s\left[2 - \cos\left(\phi + \frac{\nu_{\text{slip}}}{2}\theta\right) - \cos\left(\phi - \frac{\nu_{\text{slip}}}{2}\theta\right)\right]$$
$$+ \left[\frac{a(\bar{p})\nu_{\text{slip}}}{2}\cos\left(\phi + \frac{\nu_{\text{slip}}}{2}\theta\right) - \frac{b(\bar{p})\nu_{\text{slip}}}{2}\cos\left(\phi - \frac{\nu_{\text{slip}}}{2}\theta\right)\right]. \qquad (3.5)$$

We choose

$$a(\bar{p}) = \frac{2}{\alpha_s + 2\bar{p}} \quad \text{and} \quad b(\bar{p}) = -\frac{2}{\alpha_s - 2\bar{p}} \qquad (3.6)$$

to cancel the linear potential part of the Hamiltonian (3.1) with the slip-stacking parameter $\alpha_s$ defined in Eq. (2.4). After the cancellation of the terms linear in $\bar{p}$, we find

$$H = \frac{\nu_s \bar{p}^2}{2} + 2\nu_s\left[\frac{\cos(\phi + \frac{\nu_{\text{slip}}}{2}\theta)}{\alpha_s + 2\bar{p}} - \frac{\cos(\phi - \frac{\nu_{\text{slip}}}{2}\theta)}{\alpha_s - 2\bar{p}}\right]^2$$
$$= \frac{\nu_s \bar{p}^2}{2} + 2\nu_s\left[\frac{\alpha_s^2 + 4\bar{p}^2}{(\alpha_s^2 - 4\bar{p}^2)^2} - \frac{\cos 2\phi + \cos\nu_{\text{slip}}\theta}{\alpha_s^2 - 4\bar{p}^2} + \frac{\cos(2\phi + \nu_{\text{slip}}\theta)}{2(\alpha_s + 2\bar{p})^2} + \frac{\cos(2\phi - \nu_{\text{slip}}\theta)}{2(\alpha_s - 2\bar{p})^2}\right], \qquad (3.7)$$

where $\phi$ is related to $\bar{\phi}$ through Eq. (3.3). The second-order canonical perturbation to the two-rf system focuses the dynamics at the intersection region between the two buckets. All the fundamental-harmonic dynamics are embedded in the transformed canonical momentum $\bar{p}$. In other words, if higher harmonics are ignored $\bar{p}$ becomes a constant of motion. The square-bracketed part of the Hamiltonian in Eq. (3.7) describes the leftover phase-space structure in between the upper and lower buckets, or the overlapping resonances of these two rf buckets.

At the phase-space region near $\bar{p} = 0$, the Hamiltonian can be simplified to

$$H \approx \frac{\nu_s \bar{p}^2}{2} + \frac{2\nu_s}{\alpha_s^2}\left[1 - \cos 2\phi - \cos \nu_{\text{slip}}\theta \right.$$
$$\left. + \frac{\cos(2\phi - \nu_{\text{slip}}\theta)}{2} + \frac{\cos(2\phi + \nu_{\text{slip}}\theta)}{2}\right] + \mathcal{O}(\bar{p}\alpha_s^{-3}).$$
$$(3.8)$$

The term $\cos \nu_{\text{slip}}\theta$ is purely oscillatory and is independent of the canonical variable; it can be ignored. We can also neglect these two second-order "time-dependent" terms in Eq. (3.8) that arise from the second-order perturbation of the upper to lower buckets.

The stationary term $\cos 2\phi$ represents a second-order resonance induced by the overlapping upper and lower rf buckets. The key problem of the slip-stacking is that the upper and lower buckets can generate second-order buckets in between these two first-order buckets with a concatenated potential $\frac{2\nu_s}{\alpha_s^2}(1 - \cos 2\phi)$. This introduces second-order resonance islands and chaos in the overlapping regions of original rf buckets, and thus reduces the stable area of the original buckets for the injected beam. The loss of injected beams can become severe. If one can remove the second-order resonance islands generated by these two bucket series, one can minimize particle loss in the slip-stacking process. To this end, we add a second-harmonic rf cavity to cancel the second-order resonance and modify the *initial* Hamiltonian from Eq. (3.1) to

$$H = \frac{\nu_s p^2}{2} + \nu_s\left[2 - \cos\left(\phi - \frac{\nu_{\text{slip}}}{2}\theta\right) \right.$$
$$\left. - \cos\left(\phi + \frac{\nu_{\text{slip}}}{2}\theta\right) + \frac{\lambda}{2}(1 - \cos 2\phi)\right], \quad (3.9)$$

where

$$\lambda = -\frac{4}{\alpha_s^2} \quad (3.10)$$

corresponds to the strength of the compensating second-order rf harmonic potential intended to cancel the second-harmonic voltage produced by the fundamental rf cavities.

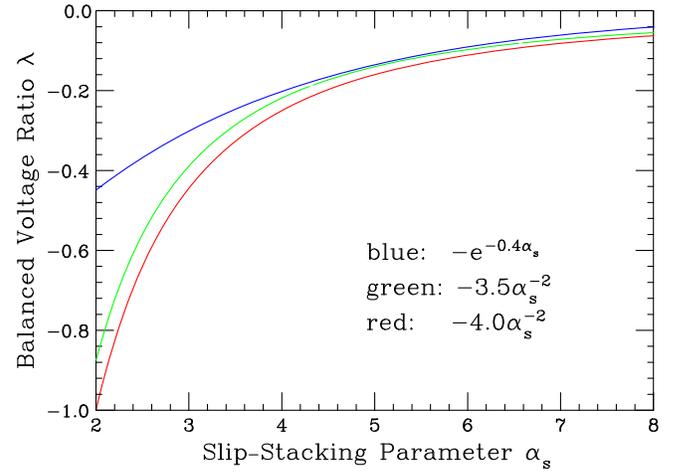

FIG. 1. Comparison of derived compensation second-harmonic ratio $\lambda$ with empirical values obtained by Eldred through simulations. Our determination is shown in red.

Here, $\frac{1}{2}\lambda$ represents the ratio of the second-harmonic rf voltage to the fundamental rf voltage [7]. The compensation by second-harmonic potential has been proposed by Eldred [5], who, through numerical tracking, determined the *optimal* amount of compensation or the value of $\lambda$ by the exhibition of the maximum stable phase-space area. His numerical fitting reveals two empirical formulas: $\lambda = -e^{-0.4\alpha_s}$ and $\lambda = -3.5\alpha_s^{-2}$ depicted in Fig. 1. For comparison, our determination is shown in red.

In short, the canonical transformation solves the Hamiltonian up to the first-order rf harmonic leaving behind all higher harmonics. Further canonical transformations can also be performed to solve for the next rf harmonic successively. However, it is normally sufficient to stop at the second-order canonical perturbation to tackle these important resonances.

## IV. NUMERICAL TRACKING

The Fermilab accelerator complex for the slip-stacking is composed of the fast-cycling Booster and the Recycler Ring. The rf buckets at the Booster and the Recycler are both generated by rf systems of the same frequency $\sim$52.8 MHz. The Booster accommodates $h_B = 84$ rf buckets and has a cycle rate of $f_B = 15$ Hz. The slip-stacking frequency is therefore $f_{\text{slip}} = f_B h_B = 1260$ Hz. The Recycler has a revolution period of $T_0 = 11.13~\mu s$ or a revolution frequency of $f_0 = 89.85$ kHz. Thus the slip-stacking tune is $\nu_{\text{slip}} = f_{\text{slip}}/f_0 = 0.0140$. The Recycler operates at the fixed particle kinetic energy of 8 GeV with a phase-slip factor $\eta = -0.00869$. At the rf voltage $V_{\text{rf}} = 60$ kV, for example, the small-amplitude synchrotron tune of the Recycler, is $\nu_s = 0.00235$ and the slip-stacking parameter is $\alpha_s = \nu_{\text{slip}}/\nu_s = 6.0$. Figure 2 shows the slip-stacking parameter $\alpha_s$ versus the rf voltage $V_{\text{rf}}$ for the Recycler.

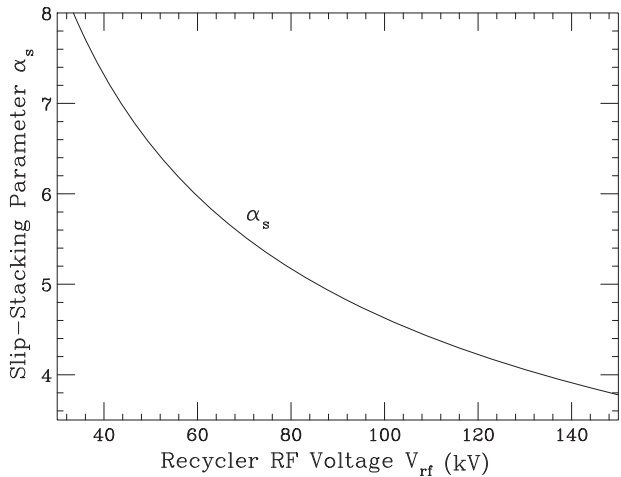

FIG. 2. The slip-stacking parameter $\alpha_s$ versus the Recycler rf voltage $V_{rf}$.

Numerical simulations of the Hamiltonian dynamics [Eq. (3.9)] are performed in the normalized phase-space coordinates. In theory, no machine parameters are needed. The above quoted machine parameters just give us an idea of the parameter space to be considered in a realistic accelerator complex. The only machine parameter employed in the tracking is the synchrotron tune, which sets the pace for the phase and off-momentum coordinates of a particle in the longitudinal phase space per revolution.

The number of revolutions in the turn-by-turn tracking or the slip-stacking time is a machine-dependent number. The injection of six booster batches takes $6/f_B = 0.40$ s, or $6f_0/f_B = 3594$ Recycler turns. This is the number of turns each tracking was performed. In the tracking with a phase modulation or a voltage modulation, stroboscopic frames are taken every modulation cycle, to ensure that the phase-space diagram remains stationary or nonrotating. Here, synchrotron oscillations in the Recycler are modulated by the slip-stacking tune $\nu_{slip}$. The stroboscopic frames are therefore taken every slip-stacking cycle or every $N_{slip}$ turns, with $N_{slip} = 1/\nu_{slip} = f_0/f_{slip}$ exactly. Unfortunately, $1/\nu_{slip} = 71.28$ is not a whole number. To facilitate the viewing of the tracking results, we modify the slip-stacking cycle to exactly $N_{slip} = 71$. Since the slip-stacking frequency is fixed at $f_{slip} = 1260$ Hz, the revolution frequency of the Recycler Ring is modified accordingly to $f_0 = N_{slip}f_{slip} = 89.46$ kHz (instead of the commonly quoted revolution frequency of 89.85 kHz). The stroboscopic phase-space plot is called the Poincaré map. We have studied tracking at various values of the slip-stacking parameter $\alpha_s$, which is an input to our tracking code. Below, we report tracking results at $\alpha_s = 3.5, 4.1$, and 6.0. All of these Poincaré maps have the same initial phase-space points [8].

First, we study the situation of $\alpha_s = 4.1$ (or $V_{rf} = 127.3$ kV for the Fermilab Recycler). Without second-harmonic rf compensation, i.e. $\lambda = 0$, the phase-space structure is shown in the top plot of Fig. 3, where there are many structure resonances embedded in a large chaotic region between the upper and lower rf buckets. The lower plot shows the phase-space structure with the second-harmonic rf compensation. The compensation causes some structure resonances in the buckets to be bounded by invariant tori and the stable bucket area increases dramatically to accommodate a larger injection bunch.

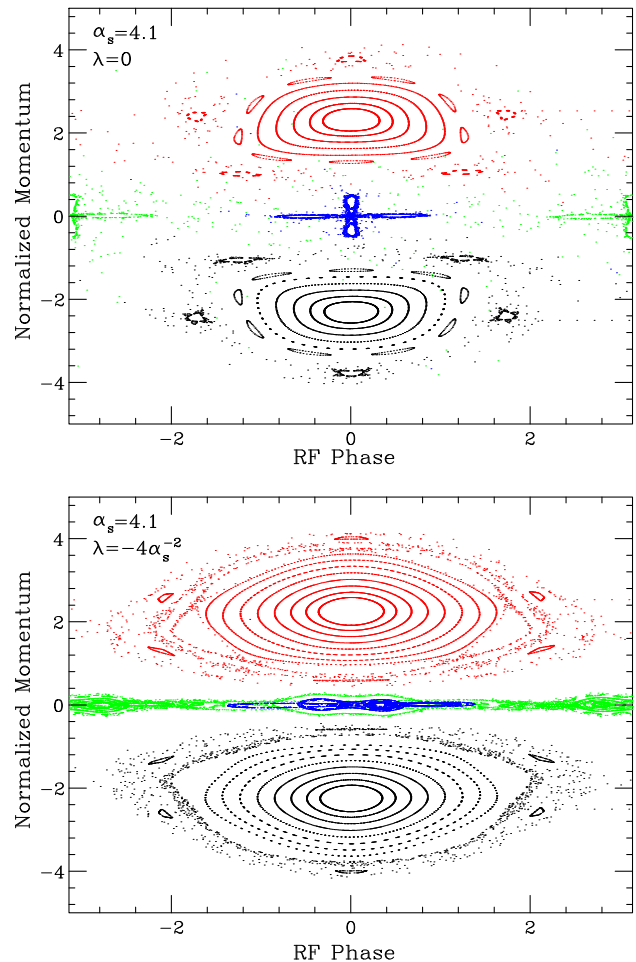

FIG. 3. Top: phase-space structure with slip-stacking parameter $\alpha_s = 4.1$ without second-harmonic rf compensation, i.e. $\lambda = 0$. Bottom: phase-space structure when the second-harmonic rf compensation is turned on, or $\lambda = -4/\alpha_s^2 = -0.238$. The interaction of two primary buckets is so strong that the phase space between them becomes chaotic. Only the cores of the fifth-order resonance islands are stable in the top plot. Once the compensation rf is included, the chaotic region becomes smaller. There are stable tori outside the fifth-order resonance islands and the stable bucket area increases, i.e., a larger injection bunch area can be accepted.

In order to depict the upper and lower rf buckets together with the central fourth-order structure in one plot, particles in the upper and lower buckets are made to slip backward artificially by the phase $\pm\pi\nu_{slip}$ per revolution turn. This back-slipping process is just to facilitate plotting and is not incorporated in the turn-by-turn tracking. This corresponds to the Poincaré maps in the respective rotating frames.

Without second-harmonic rf compensation, the cores of the fifth-order resonance islands are embedded in the chaotic sea (upper plot). The fifth-order resonance is one of the fundamental resonances generated by the modulation of the synchrotron oscillations by the slip-stacking frequency. Its occurrence comes about because $\nu_{slip}/\nu_s = \alpha_s > 4$. It is the choice of $\alpha_s = 4.1$ (just above 4) that produces the fifth-order resonance islands nearer to the separatrix of each rf bucket than the center (see Sec. IV A below). Increasing $\alpha_s$, these fifth-order resonance islands moves toward the center of the bucket [9].

When the second-harmonic rf is turned on at $\lambda = -4/\alpha_s^2 = -0.238$, the second-order resonances between the upper and lower buckets at $\phi = 0$ and $\pi$ are simultaneously compensated. Reducing the mutual interactions between the two bucket series, the fifth-order resonance islands are now enclosed inside stable tori (lower plot). The stable bucket area is much enlarged and the acceptance of the slip-stacking beam has been substantially increased.

We next study the situation when the slip-stacking parameter is $\alpha_s = 6.0$, the typical Fermilab slip-stacking operational condition at $V_{rf} = 60$ kV. The results without and with second-harmonic rf compensation are shown in the upper and lower plots of Fig. 4. Here, the first fundamental parametric resonance generated by the slip-stacking modulation is the seventh-order resonance and the resonance islands are very far away from the centers of the buckets.

The second-harmonic rf compensation $\lambda = -4/\alpha_s^2 = -0.111$ appears to change little the upper and lower buckets. This is understandable because these two buckets are well separated in the *normalized* fractional-momentum coordinate $p$, or the buckets are much narrower in the *unnormalized* fractional-momentum-spread coordinate $\delta$. The mutual influence between the two buckets becomes much smaller, and so is the modulation at the slip-stacking tune $\nu_{slip}$. This explains why the seventh-order resonance is less pronounced. Although the compensation term does reduce the resonances between the upper and lower beams at $p = 0$ and $\phi = 0, \pi$, the compensation does not increase the bucket area for the stacking beams by much. There is *no* significant reduction in the chaos near the separatrices of the upper and lower buckets, because they arise from overlapping parametric resonances.

In case the bunch area of the RCS beam is too large, we may need a larger slip-stacking rf voltage $V_{rf}$ leading to the employment of a slip-stacking parameter $\alpha_s < 4$. In an extreme case, we examine the effect of resonance compensation for $\alpha_s = 3.5$, or $V_{rf} = 174.7$ kV at the Fermilab Recycler. The top plot of Fig. 5 shows the Poincaré map of slip-stacking buckets without correction. The stable phase-space area of the bunch is the small diamondlike region of the bucket (top plot), the signature of a very strong fourth-order resonance. Because the slip-stacking parameter is 3.5, the fourth-order resonance islands reside about midway between the center of the bucket and its separatrix. However, they are not visible, because the interaction between the upper and lower buckets is so strong that all resonance islands from the interaction are destroyed and the chaotic region covers nearly all the phase space between the upper and lower buckets.

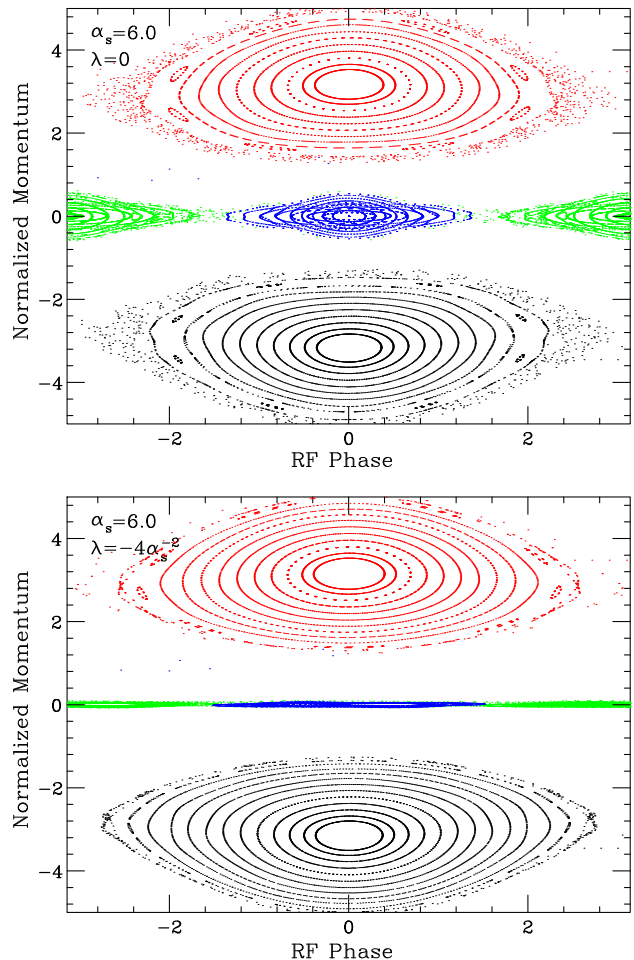

FIG. 4. Phase-space structure with slip-stacking parameter $\alpha_s = 6.0$ is shown in the top plot with no second-harmonic rf compensation or $\lambda = 0$, and in the lower plot with second-harmonic rf compensation or $\lambda = -4/\alpha_s^2 = -0.111$. It is clear that the compensation reduces the chaotic region between the upper and lower buckets with, however, only small increase in the phase-space areas of the slip-stacking buckets.

When the second-harmonic compensation cavity with $\lambda = -4/\alpha_s^2$ is turned on, the stable phase-space area increases considerably (bottom plot). The buckets restore to their pendulumlike or elliptical shapes. The strong fourth-order resonance islands are now visible in the upper and lower slip-stacking buckets. However, they are still embedded in the chaotic sea. The compensation term helps the perturbation between the buckets but cannot change the resonances within each bucket. The stable part of the rf buckets may be still inadequate to accept a sizable bunch from the Booster. It is therefore advisable to choose the slip-stacking parameter at $\alpha_s > 4$ in order to avoid beam loss, even with the compensation term turned on.

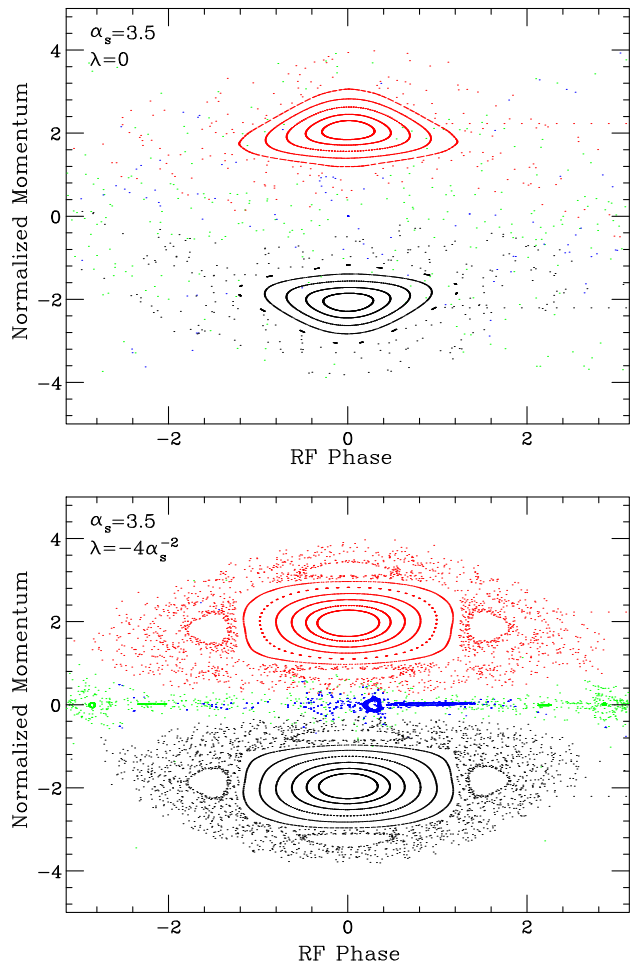

FIG. 5. Phase-space structure with slip-stacking parameter $\alpha_s = 3.5$ is shown in the top plot with no second-harmonic rf compensation or $\lambda = 0$, and in the lower plot with second-harmonic rf compensation or $\lambda = -4/\alpha_s^2 = -0.327$. The compensation restores some of the structure resonances, but they are still embedded in the chaotic sea. The compensation does not cancel resonances associated with each slip-stacking bucket.

### A. Parametric resonances

For beam particles in one of the buckets, the other slip-stacking rf system produces a time-dependent modulation at the tune of $\nu_{\text{slip}} = \alpha_s \nu_s$, which is a combination of phase and voltage modulations. If the modulation tune is equal to an integer multiple of the particle tune, resonance occurs. The synchrotron tune of a particle in the bucket depends on its synchrotron amplitude [6]; i.e., the synchrotron tune is $\nu_s$ at small amplitude and decreases to zero at the separatrix of the synchrotron phase space, as shown in Fig. 6. Resonances will occur at different phase-space locations as the slip-tune $\alpha_s \nu_s$ changes.

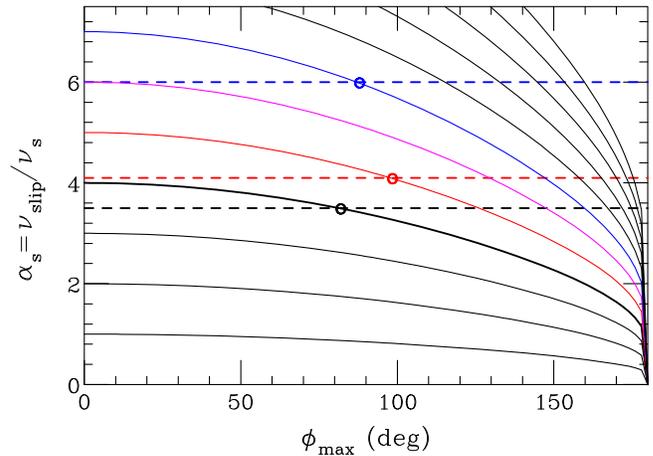

FIG. 6. The synchrotron tune and its harmonics of a stationary rf system versus $\phi_{\text{max}}$, the maximum phase amplitude of synchrotron oscillation. The horizontal dashed lines are modulation tunes at various slip-stacking parameters. When they cut through an $n$th harmonic of the synchrotron tune, the $n$th-order resonance will occur at that phase-space amplitude. These are the resonances visible in Figs. 3–5.

If the modulation (slip-stacking) tune cuts through the $n$th harmonic of the synchrotron tune, the $n$th-order resonance, called the $n : 1$ parametric resonance, will appear at the corresponding phase-space location. For example, the horizontal red dashed line in Fig. 6 corresponding to $\alpha_s = 4.1$ cuts through the fifth harmonic of the synchrotron tune to produce a fifth-order resonance at the maximum rf phase amplitude $\phi_{\text{max}} \sim 95°$. Two strong resonances can also concatenate into a second-order resonance, for example, the $4 : 1$ and $5 : 1$ resonances can interact to produce a $9 : 2$ resonance at the phase space in between these two first-order resonances, evidently shown at the top plot of Fig. 3. The size of the resonance islands depend on the resonance strength and the slope of the $n$-harmonic synchrotron tune versus amplitude [6].

In the slip-stacking Hamiltonian, the resonances created by the interaction between the upper and lower rf buckets are called the second-order and higher-order resonances. For example, the second term of the Hamiltonian of Eq. (3.8) is the second-order resonance. When the second-order resonance and parametric resonance overlap, resonance islands of both types are partially or totally destroyed, leaving behind a sea of chaos. This explains why the fourth-order resonances are not visible in the upper plot of Fig. 5. In the presence of second-harmonic rf compensation, the second-order resonance is mostly canceled, and their interaction with the parametric resonances reduced. This explains why the fourth-order resonances in the lower plot of Fig. 5 are restored and their resonance islands become visible. Unfortunately, these fourth-order resonance islands are still outside any stable tori. They are still surrounded by a chaotic sea, which is the result of strong overlapping parametric resonances generated by time modulation at $\nu_{slip} = \alpha_s \nu_s$, and cannot be canceled by the compensating second-harmonic rf system.

## V. CHOICE OF SLIP-STACKING PARAMETER $\alpha_s$

In this section, we apply the slip-stacking Hamiltonian to the Fermilab accelerator complex. The slip-stacking frequency of the Recycler is fixed at $f_{slip} = h_B f_B = 1260\,\text{Hz}$, where $h_B = 84$ is the rf harmonic number of the Booster and $f_B = 15\,\text{Hz}$ is its repetition or cycle rate. The rf harmonic number for the Recycler is $h = 588$. The frequencies of the two rf systems in the Recycler are separated by $f_{slip}$, and the centers of the upper and lower bucket series are separated by the fractional off-momentum $\Delta\delta_{sep} = \nu_{slip}/(h|\eta|) = 0.00276$, where $\eta = -0.00869$ is phase-slip factor of the Recycler Ring. Given a slip-stacking parameter $\alpha_s$, the necessary Recycler rf voltage $V_{rf}$ is

$$V_{rf} = \frac{2\pi\beta^2\nu_s^2 E}{eh|\eta|} = \frac{2\pi\beta^2 E\nu_{slip}^2}{eh|\eta|}\alpha_s^{-2}, \qquad (5.1)$$

where $E = 8.938\,\text{GeV}$ for the total nominal energy of the particle beam at slip-stacking.

Since the separation of the two rf bucket series is, in terms of the normalized fractional momentum spread, $p_{sep} = \alpha_s$, it is more transparent to use the slip-stacking parameter. But, how does one choose the slip-stacking parameter $\alpha_s$ in operation? Note that the choice of $\alpha_s$ does not change the unperturbed bunch area of the captured combined slip-stacked beam by much.

If one chooses $\alpha_s \gg 4$ so that two rf buckets do not overlap, the rf voltage $V_{rf}$ will be small. The unperturbed bucket area (in $\phi$-$\delta$) may be too small to accept the injected beam, resulting in possible beam loss. If one chooses $\alpha < 4$, the rf voltage $V_{rf}$ will be large, and so is the area of each rf unperturbed bucket. Since two overlapping buckets can produce strong resonances and chaos, the actual stable bucket area may become too small for beam injection.

Typically, one would choose $\alpha_s \gtrsim 4$ (or $V_{rf} < 134\,\text{kV}$) to avoid resonances and chaotic regions created by the overlapping of the upper and lower series of rf buckets. The left plot of Fig. 7 shows the unperturbed bucket area (red line), the unperturbed half-rf bucket height $\Delta E_{bucket}$ (black line) of a single rf system versus the Recycler rf voltage $V_{rf}$. The data of stable bucket half-height are obtained from our simulations with (blue) and without (black) second-order harmonic correction. The right plot shows the half-bunch width of the Booster ring at extraction for, respectively, phase-space area $\mathcal{A} = 0.1,\,0.15,$ and $0.2\,\text{eVs}$ versus the Booster rf voltage at extraction. The typical phase-space area of a Booster bunch at injection is about $\mathcal{A} = 0.1\,\text{eVs}$ [10]. Although the Recycler has enough bucket area for the Booster beam, there is a mismatch in bunch shape between the two rings. At 60 to 70 kV Recycler rf voltage, the bunch aspect ratio is 1.4 to 1.5 MeV/ns, while at the 350 kV Booster extraction rf voltage, the aspect ratio of the bunch is 5.5 MeV/ns. Without suitable bunch rotation, the height of the stable part of the slip-stacking (or perturbed) bucket is not sufficient for the Booster beam.

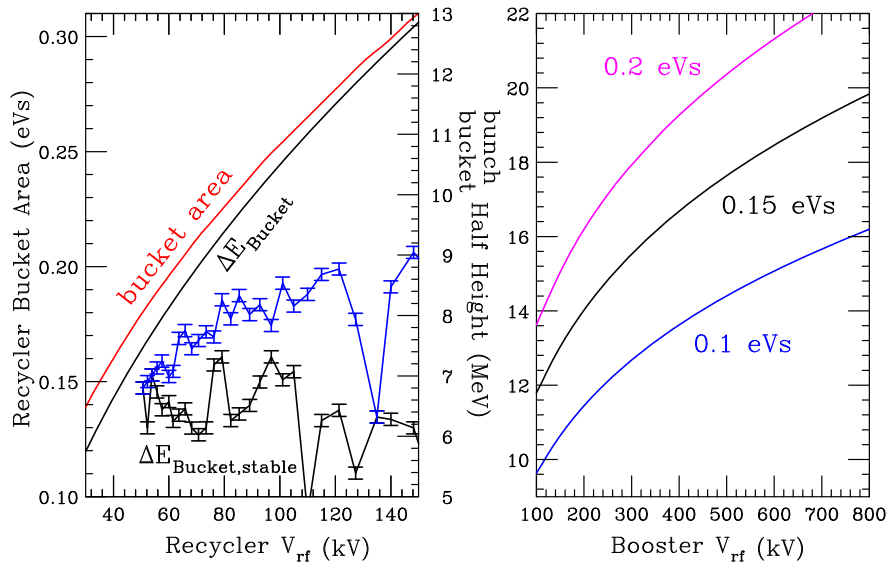

FIG. 7. Left: bucket area (red line) and bucket height (black line) versus rf voltage $V_{rf}$ for the Recycler. The typical operational Recycler rf voltage is 60 kV. Data points are obtained from numerical simulation without (black) and with (blue) second-harmonic rf compensation, where the dip at about 90 kV arises from the fifth-order resonance inside the slip-stacking bucket. The fourth-order resonance occurs around 150 kV Recycler rf voltage. Right: bunch height of the Booster beam at bunch areas 0.1, 0.15, and 0.2 eVs versus Booster rf voltage at extraction with the Booster rf sum of about 350 kV at extraction [10].

The stable half-bucket height in the left plot of Fig. 7 shows dips of the available bucket height. These dips are located at $\alpha_s = 5.5, 5.1, 4.4, 4.1$ or $V_{rf} = 70, 82, 110$, and 127 kV, respectively. They are associated with the interaction between the resonances generated by the overlapping bucket series and the parametric resonances generated in each rf bucket by the slip-stacking modulation. With the second harmonic compensation, all these dips disappear. In the Recycler, the available stable half-bucket height of the slip-stacking system provides just about 7.0 MeV for the injected beam (see the left plot of Fig. 7 at $V_{rf} = 60$ kV or $\alpha_s = 6.0$). With compensation, the half-bucket height may be able to reach 8.5 MeV at $V_{rf} = 70$–$80$ kV with $\alpha_s \sim 5.5$ to 5.2.

The Booster bunch area does increase during the Booster cycle because of various reasons, like space-charge effects, transition crossing, horizontal and vertical coupling, etc. [10]. Eldred [5] reported a measured Booster bunch area of 0.17 eVs near extraction. If bunch rotation can be implemented to the Booster to achieve a bunch aspect ratio of 1.45 MeV/ns, the bunch half-height will be reduced to 8.86 MeV (or $\Delta\delta_b = 1.0 \times 10^{-3}$). Then with a proper compensating second-harmonic rf cavity, the Recycler rf voltage of $V_{rf} = 70$ to 80 kV (corresponding $\alpha_s = 5.5$ to 5.2) should be able to provide sufficient bucket height for the Booster bunch accommodation.

Our simulations show that the second-harmonic compensation is a beneficial for $\alpha_s < 6.0$. A proper second-harmonic compensation correction will enlarge the stable part of the slip-stacking buckets for the injection beam. The compensation rf cavity, however, cannot compensate the intrinsic parametric resonances within each slip-stacking bucket, because those resonances are generated by the modulation of the synchrotron oscillations at the modulation tune of $\nu_{slip} = \alpha_s \nu_s$. The key to achieving efficient slip-stacking is to optimize bunch-phase-space matching between the RCS and the slip-stacking ring.

## VI. CONCLUSION

In conclusion, we have employed the Hamiltonian dynamics to study the slip-stacking dynamics, and we use numerical simulations to verify the Hamiltonian dynamics. We apply the slip-stacking Hamiltonian to the Fermilab Booster-Recycler complex. We have derived the optimal range of the slip-stacking parameter and its dependence on the phase-space area at the Booster extraction. To improve the slip-stacking efficiency, it is important to minimize the phase-space area of the Booster extracted beam, but the key problem is the aspect-ratio matching of the RCS to the slip-stacking ring.

The final slip-stacked phase-space area depends essentially on the Booster repetition rate and the properties of the Recycler, for example, the rf harmonic numbers and the phase-slip factor [see Eq. (2.3)]. Once these parameters are fixed, the phase space of the stacked beam is fixed. The slip-stacking parameter does not affect the captured stacked-beam area by much. If the phase-space area of the injected beam is small, one can use $\alpha_s \gg 4$ to minimize the interaction of the slip-stacking buckets. However, if the phase-space area of the injection beam is large, one can use the compensation second-harmonic rf cavity to minimize the interaction between the slip-stacking buckets at $\alpha_s \lesssim 5.5$. The overlapping of the upper and lower rf bucket series produces the resonance phase-space structure in between the buckets, while the modulation of the synchrotron oscillations by the slip-stacking cycles produces the intrinsic parametric resonances inside the rf buckets [6,9]. When these two types of resonances overlap, chaos results. With the compensation rf cavity of Eq. (3.9), resonances in between the two buckets are removed leaving intact the parametric resonances associated with the buckets. The compensation second-harmonic rf cavity can only cancel to a certain extent the phase-space structure between the bucket series. It cannot compensate the intrinsic parametric resonances inside each slip-stacking bucket. If parametric resonances are bounded by invariant tori, particles inside these tori will be stable. Thus the compensating rf system would be useful in recovering the stable bucket area in some cases. Figure 7 of our simulations can be experimentally tested at Fermilab on the stable bucket height for the slip-stacking beams at Fermilab.

## ACKNOWLEDGMENTS

We thank J. Eldred and R. Zwaska for fruitful communications. This work is supported in part by grants from the National Science Foundation (NSF) No. PHY-1504778 and the U.S. Department of Energy under Contract No. DE-AC02-76CH030000.